\documentclass[12pt]{article}
\usepackage{epsfig}

\def\beq{\begin{equation}}
\def\eeq#1{\label{#1}\end{equation}}
\def\eeqn{\end{equation}}

\def\beqa{\begin{eqnarray}}
\def\eeqa#1{\label{#1}\end{eqnarray}}
\def\eeqan{\end{eqnarray}}

\let\bar=\overbar

\def\Dslash{\not{\hbox{\kern-4pt $D$}}}
\def\dslash{\not{\hbox{\kern-2pt $\del$}}}

\def\msb{{\bar{\ssstyle M \kern -1pt S}}}

\def\Title#1{\begin{center} {\Large {\bf #1} } \end{center}}

\begin{document}

\Title{Color-flavor locking in strange stars, strangelets, and cosmic
rays}

\bigskip\bigskip


\begin{raggedright}  

{\it Jes Madsen\index{Madsen, J.}\\
Institute of Physics and Astronomy\\
University of Aarhus\\
DK-8000 {\AA}rhus C, DENMARK}
\bigskip\bigskip
\end{raggedright}

\section{Introduction}

Three topics with relation to color superconductivity in strange quark
matter are discussed. 
1) The $r$-mode instability in strange stars,
which is consistent with the existence of ``ordinary'' strange quark
matter stars but inconsistent with strange stars in a pure color-flavor
locked state.
2) Color-flavor locked strangelets, which are more bound than normal
strangelets, and have a different charge-mass relation.
3) Estimates of the strangelet flux in cosmic rays, which is relevant
for strangelet detections in upcoming cosmic ray space experiments.

\section{$r$-mode instabilities in strange stars}

As discussed in several talks during this conference, it is not an easy
task to distinguish observationally between neutron stars and strange
stars, and it is even more difficult to tell the difference between
``ordinary'' strange stars and strange stars composed of color-flavor
locked (CFL) quark matter and/or quark matter with a two-flavor color
superconducting phase (2SC). Precise mass and radius measurements may
offer a way because strange stars are generally more compact than
neutron stars. Differences in neutrino cooling properties could be
another possibility. While some data can be interpreted as being in
favor of at least a few objects being strange stars, the situation is
still not convincingly settled \cite{madsen99,alford01b}.

Another phenomenon in compact stars has turned out to be a sensitive probe
of quark matter properties. This is the $r$-mode instability, which 
is a generic instability in all rotating compact stars in the absence of
viscous forces \cite{andersson98,friedman98,andkok01}. 
The instability involves horisontal mass-currents in the
stellar fluid. These currents couple to gravitational
wave emission (like magnetic quadropole radiation). In a rotating star
there are $r$-modes which are counter-rotating as seen from the stellar
rest frame, but forward-rotating as seen from infinity. Gravitational
wave emission taps positive energy and angular momentum from the mode,
which strengthens the mode in the stellar frame, so the mode is
inherently unstable for any rate of rotation. 

Viscosity may, however, prevent the mode from growing. In ordinary
neutron stars, the combination of shear and bulk viscosities, and in
particular the effect of ``surface rubbing'' \cite{bild00}
between the inner fluid and the
solid part of the stellar crust dampens the $r$-mode instability
significantly, leaving an interesting regime only for very high temperature
and rotation rates close to the mass-shedding limit (the so-called
Kepler limit, where the equatorial centripetal and gravitational
accelerations are equal).

``Ordinary'' strange stars may also have solid crusts of nuclear matter,
held afloat by a strong electrostatic potential at the surface of the
quark phase (the electrons are not as strongly bound as the quarks, so
they form a thin ``atmosphere'', corresponding to a strong outward
directed electrostatic potential), but the maximal crust density is much
smaller than for neutron stars, only $\approx 10^{11} {\rm g~cm}^{-3}$,
and therefore the surface rubbing is far less important.

In fact, data on pulsar rotation are consistent 
with the strange star hypothesis
if the quark matter is non-superfluid \cite{madsen98,madsen00a}, 
and if the strange stars are
either completely bare (making them invisible in x-rays), 
or have a very thick crust (to assure an internal temperature much
higher than the surface temperature; otherwise the most rapid
millisecond pulsars would be in a regime with significant spin-down due
to the $r$-mode instability, contrary to observations). 

\begin{figure}[h!tb]
\begin{center}
\epsfig{file=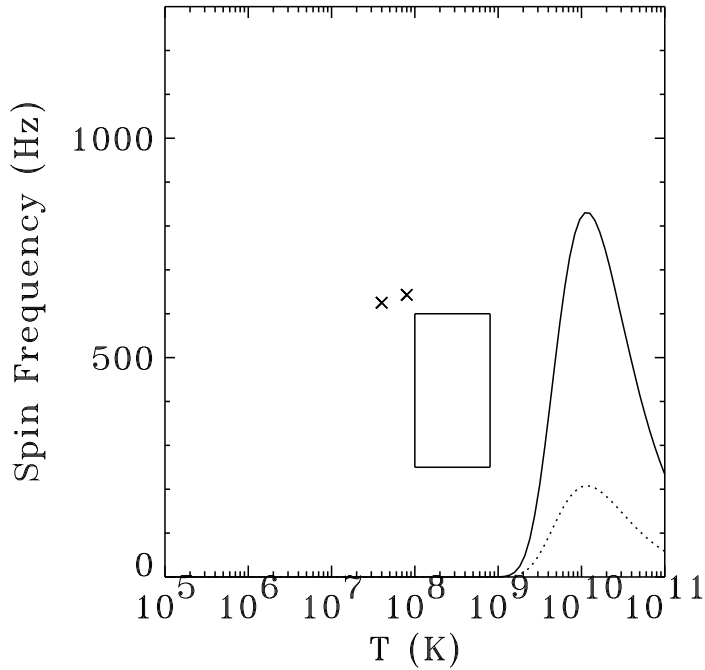,height=4.0in}
\caption{Critical rotation frequencies in Hz as a function of 
internal stellar temperature for CFL strange stars with an 
energy gap as small as $\Delta=1$~MeV.
Full curve is for $m_s=200$~MeV; dotted curve for $m_s=100$~MeV.
The box marks the positions of most low-mass x-ray binaries
(LMXB's), and the crosses are the most
rapid millisecond pulsars known (the temperatures are upper limits).
All strange stars above the curves (i.e.\ essentially all over the
diagram) would spin down on a time scale of hours due to the $r$-mode
instability, in complete contradiction to the observation of millisecond
pulsars and LMXB's. Thus CFL strange stars are ruled out.}
\label{fig:cfl}
\end{center}
\end{figure}

In contrast strange stars purely in a color-flavor locked phase are not
permitted by pulsar data \cite{madsen00a}. 
The main contribution to bulk viscosity in strange quark matter is the
weak reaction ${\rm u} + {\rm d}\leftrightarrow {\rm s} + {\rm u}$,
whereas shear viscosity is governed by strong quark-quark scattering.
If the quark pairing energy gap is $\Delta$,
the characteristic time-scales for viscous damping of $r$-modes are
exponentially increased by factors of $\exp (2\Delta /T)$ and $\exp (
\Delta /(3T))$ respectively for bulk and shear viscosity. As the
relevant temperature regime in pulsars is $T\ll 1$~MeV (except in the
first few seconds after the supernova explosion) this means that viscous
damping of $r$-modes essentially disappears for $\Delta >1$~MeV, and all
color-flavor locked strange matter pulsars would spin down within hours
in sharp contrast to observations. Figure \ref{fig:cfl} illustrates the
situation for $\Delta=1$~MeV. Values of $\Delta$ as high as 100~MeV are
often assumed; for such values the entire diagram would be $r$-mode
unstable.

\begin{figure}[h!tb]
\begin{center}
\epsfig{file=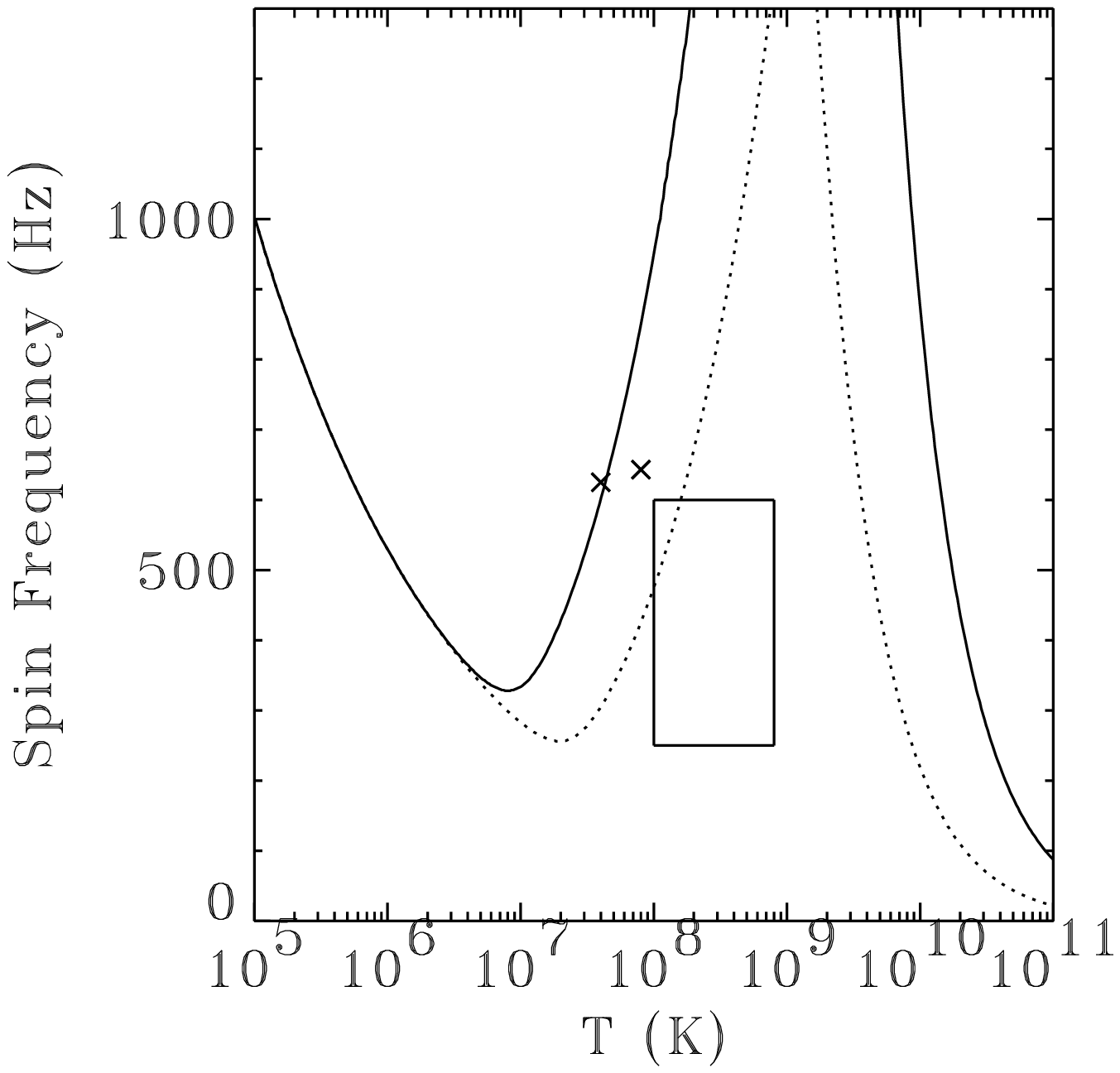,height=4.0in}
\caption{As Figure \ref{fig:cfl}, but for 2SC-stars. 
Rapid spin-down happens above
the parabola-like curves. The rapid millisecond pulsars are
uncomfortably close to the instability regime, in particular because the
temperatures are upper limits, but it would not seem appropriate to rule
out 2SC strange stars on this basis alone.}
\label{fig:2sc}
\end{center}
\end{figure}

Thus, the $r$-mode instability seems to firmly rule out that pulsars are
color-flavor locked strange stars. (In \cite{madsen00a} I showed how
electron shear viscosity and/or surface rubbing on a CFL strange star
crust might stop the spin-down at rotation periods of order 10
milliseconds at intermediate temperatures. This would still be far too
slow to be consistent with observations of pulsars and low-mass x-ray
binaries, but even these viscous mechanisms are ruled out by the
realisation, that CFL strange quark matter is charge neutral
\cite{rajwil01}, with equal
numbers of up, down, and strange quarks, and no electrons. Such a system is
without the electron ``atmosphere'' needed to sustain a crust; thus
color-flavor locked strange stars would only exist in a ``bare'' variety,
but these would not be able to
rotate at all because of the $r$-mode instability, 
and are therefore not the pulsars or low-mass x-ray binaries observed).

For strange quark matter in the 2SC phase the situation is less
conclusive. Now only some of the strong and weak reaction channels
responsible for shear and bulk viscosities are exponentially suppressed,
and the resulting viscous timescales for $r$-mode damping are increased in
a much less dramatic fashion than for the CFL-phase, by factors of 9 and
$(9/5)^{1/3}$ respectively for bulk and shear viscosity \cite{madsen00a}. 
As seen in Figure \ref{fig:2sc} it becomes
difficult to reconcile pulsar and LMXB data with 2SC-calculations, but
it would be premature to rule it out.

The discussion above focused on strange stars, i.e.\ on the assumption
of absolutely stable strange quark matter. If quark matter is only
metastable CFL and/or 2SC phases could exist in the interior of hybrid
stars, with mixed phases and ordinary hadronic matter in the outer
parts. While no explicit $r$-mode studies have been performed for such
systems, it is clear that the results will resemble those for
ordinary neutron stars. This is because $r$-modes are located mainly in
the outer parts of the star, and they are significantly damped by the
boundary condition stemming from the fact that the $r$-modes cannot
propagate in a solid crust. 

Even pure quark matter stars could have significant substructure which
could make them resemble neutron or hybrid stars in terms of $r$-mode
instabilities. For instance, if the density profile of the star is such
that only the central region is in the CFL phase, but the outer part in
2SC, then crystalline like structures like the LOFF-phase could form
\cite{alford01},
and a surface rubbing effect might appear, suppressing the $r$-modes.

Finally magnetic fields and the exact nature of the superfluid vortices
could be important. However, the conclusion that pure CFL strange stars
are ruled out by the $r$-mode studies appears to be robust.

\section{Color-flavor locked strangelets}

If quark matter is in a color-flavor locked phase, it is because this
phase has lower energy than other possible phases. In particular, this
means that metastability or even absolute stability of strange quark matter
becomes more likely than hitherto assumed, based on calculations for
``ordinary'' strange quark matter \cite{madsen99,alford01b}. 
For relevant ranges of strange quark
mass, the gain in energy per baryon for bulk strange quark matter is
roughly 100~MeV at fixed bag constant for $\Delta=100$~MeV.

This also makes it relevant to reconsider the properties of finite size
quark matter lumps, strangelets, for color superconducting strange quark
matter. A first attempt in this direction was a study of color-flavor
locked strangelets within the framework of the multiple-reflection
expansion of the MIT bag model \cite{madsen01}. In this approach the
total energy (mass) of a strangelet can be written as
\begin{equation}
E=\sum_i (\Omega_i +N_i\mu_i) +(\Omega_{{\rm pair,}V} +B)V ,
\end{equation}
where the sum is over flavors, $B$ is the bag constant, $N_i$ and
$\mu_i$ are quark number and chemical potential,
$\Omega_{{\rm pair,}V}\approx -3\Delta^2\mu^2/\pi^2$ is the binding
energy from pairing ($\mu$ is the average quark chemical potential), and
the thermodynamic potential of quark flavor $i$ is a sum of volume,
surface, and curvature terms derived from a smoothed density of states.

Apart from the pairing energy another crucial difference relative to
non-CFL strangelet calculations is the equality of all quark Fermi
momenta in CFL strange quark matter. This property, which leads to
charge neutrality in bulk without any need for electrons
\cite{rajwil01}, is due to the
fact that pairing happens between quarks of different color
and flavor, and opposite momenta $\vec p$ and $-\vec p$, so it is
energetically favorable to fill all Fermi seas to the same Fermi
momentum, $p_F$.

\begin{figure}[h]
\begin{center}
\epsfig{file=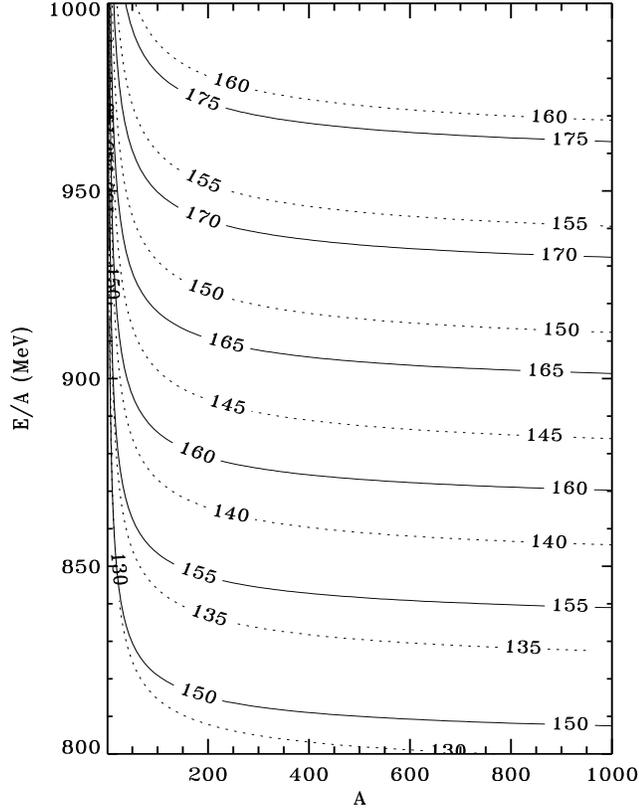,height=4.5in,width=3.5in}
\caption{Energy per baryon in MeV as a function of $A$ for ordinary strangelets
(dashed curves) and CFL strangelets (solid curves) for $B^{1/4}$ in MeV
as indicated, $m_s=150$~MeV, and $\Delta=100$~MeV.}
\label{fig:eovera}
\end{center}
\end{figure}

As illustrated in Figure \ref{fig:eovera}, 
color-flavor locked strangelets have an
energy per baryon, $E/A$, that behaves much like that of ordinary
strangelets as a function of $A$. For high $A$ a bulk value is
approached, but for low $A$ the finite-size contributions from surface
tension and curvature significantly increases $E/A$, making the system
less stable. The main difference from ordinary strangelet calculations
is the overall drop in $E/A$ due to the pairing contribution, which is
of order 100~MeV per baryon for $\Delta\approx 100$~MeV for fixed values
of $m_s$ and $B$. Since $\Omega_{{\rm pair,}V}\propto \Delta^2$, the
actual energy gain is of course quite dependent on the choice of
$\Delta$.

A significant distinction between the properties of ordinary strangelets
and CFL strangelets lies in the charge properties. They have in common a
very small charge per mass unit relative to nuclei, but the exact
relation is quite different, and this may provide a way to test
color-flavor locking experimentally if strangelets are found in
accelerator experiments or (perhaps more likely) in cosmic ray
detectors. Ordinary strangelets have (roughly)
\cite{farjaf84,berjaf87,heis93}
\begin{equation}
Z\approx 0.1 \left( m_s\over {150~{\rm MeV}} \right)^2 A; ~~~~A\ll 10^3;
\end{equation}
\begin{equation}
Z\approx 8 \left( m_s\over {150~{\rm MeV}} \right)^2 A^{1/3}; ~~~~A\gg 10^3 .
\end{equation}

In contrast, CFL strangelets are described by \cite{madsen01}
\begin{equation}
Z\approx 0.3 \left( m_s\over {150~{\rm MeV}}\right) A^{2/3} .
\end{equation}
This relation can easily be understood in terms of the charge neutrality
of bulk CFL strange quark matter \cite{rajwil01} with the added effect of the
suppression of s-quarks near the surface, which is responsible for (most
of) the surface tension of strangelets. This leads to a reduced number
of negatively charged s-quarks in the surface layer; thus a total
positive quark charge proportional to the surface area or $A^{2/3}$.

In fact, a similar effect becomes important even in ordinary
strangelets, meaning that the standard $A^{1/3}$-result breaks down at
very high $A$ \cite{madsen00b}. 
And even more important, this effect is large enough to
rule out a potential disaster scenario, where negatively charged
strangelets produced in heavy ion colliders could grow by nucleus
absorption and swallow the Earth. While ordinary strange quark matter
can be negatively charged in bulk if the one-gluon exchange $\alpha_S$
is very prominent \cite{farjaf84}, 
the added positive surface charge due to massive s-quark
suppresssion is sufficient to make the overall quark charge positive for
a large range of $A$, thus preventing any such disaster
\cite{madsen00b}.

Naturally, only first steps have been made in the effort to describe
properties of color-flavor locked strangelets. First of all, the MIT bag
model is a phenomenological approximation to strong interaction physics;
it is not QCD. Secondly, while finite-size effects were included in the free
quark energy calculations, such (unknown) higher order terms were not
taken into account in the pairing energy. This approximation 
seems warranted as long as $\Omega_{\rm pair}$ itself is a perturbation
to $\Omega_{\rm free}$. And thirdly, quark level shell effects were only
taken into account in an average sense via the smoothed density of
states described as a sum of volume, surface and curvature terms. While
this is an excellent approximation to the average strangelet 
properties \cite{madsen94}, it
misses the interesting stabilizing effects near closed shells
\cite{giljaf93,schaff97} that could
make certain baryon number states longer lived than one might expect
from a glance at Figure \ref{fig:eovera}. 
A first attempt at approaching finite size
effects in 2SC quark matter in a completely different manner is
discussed in \cite{amore01}.

\section{Strangelet flux in cosmic rays}

Two cosmic environments could in principle harbor strangelet formation. The
cosmological quark-hadron phase transition $10^{-5}$ seconds after the Big Bang,
and the high density conditions in compact supernova remnants, 
which may be strange
stars composed of quark matter rather than neutron stars. 

Cosmologically
produced strangelets were for a time believed to be natural dark 
matter candidates
\cite{witten84}. In that case a significant background of largely neutral
strangelets (quark core charge neutralized by electrons) would be moving in our
galactic halo at typical speeds of 3--400 km/sec, corresponding to the depth of
the galactic gravitational potential. Several experiments have placed limits on
the abundance of these nonrelativistic strangelets 
(sometimes called quark
nuggets), but it now seems unlikely that they could form in or survive from the
very hot ($T\approx 100$~MeV) environment in the early Universe. A similar
problem faces strangelet production in ultrarelativistic heavy ion collisions; 
it has been compared to making ice cubes in a furnace.

\subsection{Strangelet production in collisions of strange stars}

A more likely origin of cosmic ray strangelets is from collisions of binary 
compact star systems containing strange stars. If strange quark matter is
the ground state of hadronic matter at zero pressure, it will be energetically
favorable to form strange stars rather than neutron stars, and it would be
expected that all the objects normally associated with neutron stars (pulsars
and low-mass x-ray binaries) would actually be strange stars.

Several pulsars are observed in binary systems containing another compact star;
the most famous such system called PSR1913+16 delivered convincing evidence
for gravitational wave emission. 
Such binaries typically move in elliptical orbits,
spiraling closer to each other because the system loses energy by gravity wave
emission. The remaining lifetime can be estimated,
and combined with estimates of the number of these binaries in our galaxy, the
expected rate of binary collisions is of order $10^{-4}~{\rm year}^{-1}$.

Several numerical studies have been performed in the literature to follow the
late stages of inspiral in systems composed of two 
neutron stars or neutron stars
orbiting black holes or white dwarfs, especially to derive 
the gravitational wave
signatures, which are of importance for upcoming gravity wave detection
experiments. No detailed calculations have been done for systems containing
strange stars, and since there are significant differences in the equation of
state it may be dangerous to rely on existing models. Nevertheless, certain
features seem robust. While details depend on assumptions about the orbit, most
collisions seem to release a fraction of the total mass (of order
$10^{-4}$--$10^{-1}M_\odot$, where $M_\odot$ is the solar mass) in
connection with the actual collision and via tidal disruption in the late stages
of inspiral. 

No realistic estimates exist at present of the mass spectrum of quark matter
lumps released during the actual collision. Lumps of matter released during the
tidal disruption phase are expected to be very large. Balancing the tidal force
trying to disrupt the star with the surface tension force 
of strange quark matter
leads to a typical fragment baryon number of 
$A\approx 4\times 10^{38} \sigma_{20} a_{30}^3$,
where $\sigma_{20}\approx 1$ is the surface tension in 
units of 20 MeV/fm$^3$ and
$a_{30}$ is the distance between the stars in units of 30~km. 

A significant fraction of the tidally released material is 
originally trapped in 
orbits around the binary stars. The typical orbital speeds of the lumps are
here $0.1 c$, and collisions among lumps are abundant. Assuming the kinetic
energy in these collisions mainly goes to fragmentation 
of the lumps into smaller
strangelets (i.e.\ that the kinetic energy is used to the extra surface and
curvature energies necessary for forming $N$ lumps of baryon number $A/N$ from
the original baryon number $A$), it can be shown for 
typical bag model parameters
that the resulting strangelet distribution peaks at mass numbers from a few
hundred to about $10^3$. This is well within the interesting regime for 
the upcoming cosmic ray experiment Alpha Magnetic Spectrometer AMS-02 on
the International Space Station \cite{ams02} (a
prototype AMS-01 was flown on the Space Shuttle mission STS-91 in 1998).
AMS-02 is a roughly 1~m$^2$~sterad detector which will analyze the flux
of cosmic ray nuclei and particles in unprecedented detail for three
years or more following deployment in 2005. It will be sensitive to
strangelets in a wide range of mass, charge and energy \cite{amsstrange}.

\subsection{Strangelet flux at AMS-02}

Strangelet propagation in the Milky Way Galaxy
is in many ways expected to be similar
to that of ordinary cosmic ray nuclei. Except for a possible background of
slow-moving electrically neutral quark nuggets confined solely by the
gravitational potential of the Galaxy, strangelets are charged and are therefore
bound to the galactic magnetic field. They lose kinetic energy by electrostatic
interactions with the interstellar medium, and they gain energy by Fermi
acceleration in shock waves, for example from supernovae. Even if accelerated to
relativistic speeds, scatterings on impurities 
in the magnetic field makes the
motion resemble a diffusion process. The solar wind as well as the Earth's
magnetic field become important for understanding the final approach to the
detector. Also, strangelets may undergo spallation in collisions with cosmic ray
nuclei, nuclei in the interstellar medium, or other strangelets. 

Detailed studies of all of these phenomena have 
recently been started in order to
understand strangelet propagation in more detail. Much depends on the
charge-to-mass relation, but the details of propagation are not even well
understood for ordinary nuclei, so clearly some uncertainty in the expectations
for the strangelet flux at AMS is inevitable. 

In the following it will be assumed that
strangelets share two of the features found experimentally for nuclei, namely a
powerlaw energy distribution: $N(E)dE \propto E^{-2.5}$, and an average
confinement time in the galaxy of $10^7$~years. 
Assuming strangelets to move close
to the speed of light, and ascribing one baryon number, $A$, to them all, the
strangelet flux at AMS-02 would be
\begin{equation}
F= 3\times 10^{12} A^{-1} ({\rm m^2~y~sterad})^{-1}\times R_{-4} \times M_{-2}
\times V_{100}^{-1} \times t_7 \times {\rm GCfraction}
\end{equation}
where $R_{-4}$ is the number of strange star collisions in our Galaxy per $10^4$
years, $M_{-2}$ is the mass of strangelets ejected per collision in units of 
$10^{-2}M_\odot$, $V_{100}$ is the effective galactic volume in units of
$100~{\rm kpc}^3$ over which strangelets are distributed, and $t_7$ is the
average confinement time in units of $10^7$ years. 
All these factors are of order
unity if strange matter is absolutely stable, though each with significant
uncertainties.
Finally, GCfraction is the fraction of strangelets surviving the geomagnetic
cutoff. Taking this cutoff to be at rigidity 6~GeV/c, 
assuming the standard $E^{-2.5}$
powerlaw for the energy distribution with a cutoff at $\beta\equiv v/c =0.01$,
and assuming a charge-mass relation $Z=0.3A^{2/3}$ as derived for color-flavor
locked strangelets, the resulting strangelet flux at AMS-02 becomes
\begin{equation}
F= 5\times 10^5 ({\rm m^2~y~sterad})^{-1}\times R_{-4} \times M_{-2}
\times V_{100}^{-1} \times t_7 .
\end{equation}
By coincidence, this result (valid for $A<6\times 10^6$; for larger $A$ 
GCfraction equals 1 and the previous expression applies) is independent of
strangelet mass. 

As should be evident from the discussion above, there are many uncertainties
involved in the calculation of the strangelet flux at AMS-02. A systematic study
of these issues has been initiated and should significantly improve our
understanding of the strangelet production and propagation. But ultimately we
must rely on experiment. So far it is reassuring, that the simple flux estimates
above lead us to expect a very significant strangelet flux in the AMS-02
experiment.

A discovery of strangelets would of course be a very significant
achievement in itself. Getting data on the charge-to-mass relation may
even allow an experimental test of color-flavor locking in quark matter.

\bigskip
This work was supported in part by The Danish Natural Science Research
Council, and The Theoretical Astrophysics Center under the Danish
National Research Foundation. I thank Jack Sandweiss and Dick Majka for
discussions and collaboration on strangelet detection with AMS-02.

\def\Discussion{
\setlength{\parskip}{0.3cm}\setlength{\parindent}{0.0cm}
     \bigskip\bigskip      {\Large {\bf Discussion}} \bigskip}
\def\speaker#1{{\bf #1:}\ }
\def\endDiscussion{}

\Discussion

\speaker{F. Weber (Notre Dame)} In your flux
determination, did you assume that all neutron stars are in fact strange
stars or that there exist two separate families of compact stars?

\speaker{Madsen} All compact stars were assumed to be strange stars.
Coexistence of two separate families is very unlikely if strange matter
is the ground state at zero external pressure, since the Galaxy would be
``polluted'' by quark lumps from binary pulsar collisions. These lumps
would trigger transition to strange stars in supernova cores.

\speaker{F. Sannino (Nordita)} How do you disentangle (CFL) or strange
droplets experimentally? What is the signature?

\speaker{Madsen} Strangelets have very low charge for a given baryon
number compared to nuclei. CFL-strangelets have a charge-mass relation
that differs from ``ordinary'' strangelets.

\speaker{J.E. Horvath (Sao Paulo)} A low charge-to-mass ratio for strangelets
would not allow substantial acceleration in supernova shocks, thus the
spectrum (and also the confinement time) is not necessarily the one
measured in cosmic rays. How does this change the estimates of the rates
in the expected range of center-of-mass energies?

\speaker{Madsen} Clearly the propagation in the Galaxy requires further
study, and we are trying to address the relevant mechanisms in detail. 
One should keep in mind, though, that these issues are not fully
understood even for ordinary cosmic ray nuclei.

\endDiscussion
 
\end{document}